\documentclass[aps,prd,preprint,tightening,showpacs]{revtex4}
\usepackage{amsmath,amssymb,amsthm,graphicx}
\usepackage[mathscr]{eucal}
\usepackage[colorlinks,linkcolor=blue,
anchorcolor=blue,citecolor=green]{hyperref}
\usepackage[dvipsnames,usenames]{color}
\newcommand{\no}{\nonumber}
\newcommand{\be}{\begin{equation}}
\newcommand{\ee}{\end{equation}}
\newcommand{\ba}{\begin{eqnarray}}
\newcommand{\ea}{\end{eqnarray}}
\usepackage{epsf,epsfig,graphics}
\usepackage{verbatim,color,ulem}
\begin{document}
\title{Equatorial light bending around Kerr-Newman black holes}
\author{You-Wei Hsiao}
\email{hsiao.phys@gapp.nthu.edu.tw}
\author{Da-Shin Lee}
\email{dslee@gms.ndhu.edu.tw}
\author{Chi-Yong Lin}
\email{lcyong@gms.ndhu.edu.tw}
\affiliation{
Department of Physics, National Dong Hwa University, Hualien, Taiwan, Republic of China}
\date{\today}

\begin{abstract}
We study the deflection angle of a light ray as it traverses on the equatorial plane of a charged spinning black hole.
We provide
detailed analysis of  the light ray's trajectory, and derive the closed-form expression  of the deflection angle due to the black hole
in terms of elliptic integrals.
 In particular,  the geodesic equation of the light ray along the radial direction can be used to define an appropriate ``effective
 potential".
 The nonzero charge of the  black {hole} shows stronger repulsive effects to prevent  light rays from falling into the black hole as
 compared with the Kerr case. As a result, the radius of the innermost circular motion of light rays with the critical impact
 parameter
  decreases as charge $Q$ of the black hole increases for both direct and retrograde motions.
Additionally,  the deflection angle decreases when $Q$ increases with the fixed impact parameter. These results will have a direct
consequence on constructing the apparent shape of a charged rotating black hole.
\end{abstract}

\pacs{04.70.-s, 04.70.Bw, 04.80.Cc}

\maketitle

\section{Introduction}
General relativity  provides a unified description of gravity as a geometric property of spacetime \cite{MIS,HAR}. In particular, the
presence of  matter and radiation  with  energy and momentum
can curve spacetime, and the light ray's trajectory will be deflected as a chief effect \cite{CHAS}.
One of the very important consequences of general relativity is the bending of a light ray in the presence of a gravitational field.
In the light of the first image of  the black hole  captured by the Event Horizon Telescope \cite{EHT1,EHT2,EHT3}, these rays will yield
the apparent shadow of the black hole for an observer in the asymptotic region, and
 the understanding of the shadows becomes very important for measuring the properties of astrophysical black holes.

Light deflection in weak gravitational field of Schwarzschild black holes was known in 1919, and served as
the starting point to develop gravitational lensing theory.
Nevertheless, light deflection in the strong gravitational field of  Schwarzschild black holes was not studied until several decades ago  by
Darwin \cite{DAR}.  It was then reexamined  in \cite{LUM,OHA,NEM,BOZ}, and
%\textcolor{blue}
{extended to the Reissner-Nordstrom
spacetime or  charged black holes \cite{EIR,SER,KEE,BHA},} and to any spherically symmetric black holes \cite{BOZ1}. Black hole lenses
were also explored  numerically by \cite{VIR1,VIR2,VIR3}.
Kerr black hole lenses were analyzed in \cite{VAZ,BOZ2,BOZ3,BOZ4,KRA,AAZ1,AAZ2}, which, in particular, found
rotating black hole apparent shapes or shadows with an optical deformation   rather than being circles as in the case of
nonrotating ones  \cite{TAK,ZAK1,ZAK2,HIO1,BAM,HIO2,AMA}.
The bending angle of light rays due to Kerr black holes on the equatorial plane was studied analytically in \cite{SVI1,SVI2,BEA,BAR}
using the null geodesic equations.
 The deflections produced in the presence of a rotating black hole  explicitly depend on the direction of motion of the light relative
 to
 the spin direction of the black hole.
In particular, the authors of \cite{SVI1} derived the closed-form expression of  the equatorial light deflection angle
in terms of elliptic integrals. However, the strong gravitation field gives rise to
the large bending of light rays near a black hole. The bending angle can be larger than $2\pi$, showing the possibility that
light rays might go around the center of the black hole several times before reaching the observer.
Apart from a primary image, a theoretically infinite  sequence of images, which we term  relativistic images, might be formed, and are
usually greatly demagnified.
The closed-form expression of the light bending angle  in an exact result~\cite{SVI1} or in some sort of asymptotic approximation might
be of great help to study these images ~\cite{BEA,BAR}, although the observation of relativistic images is a very difficult task.

Another known asymptotically flat and stationary solution of the Einstein-Maxwell field equations in general relativity
is the Kerr-Newman metric, a generalization of the Kerr metric, which describes spacetime in the exterior of a rotating charged black
hole.
Apart from
gravitation fields, both electric and  magnetic fields exist intrinsically from the black hole.
Although one might not expect that astrophysical black holes have a {large residue} electric charge,
{some} accretion scenarios were {proposed to investigate} the possibility of the spinning charged back holes~\cite{Damo}.
It is then still of great interest to extend the {previous} studies  to a Kerr-Newman black hole  \cite{DAD,VRI4,STU,CHAR}.
The central thread of this paper is to try to achieve the exact expression of the light deflection angle by a Kerr-Newman black hole in
the equatorial plane, an extension of the work in \cite{SVI1} for the Kerr black hole case.
In Sec.II, we focus on circular trajectories of light rays  arriving from and returning to spatial infinity.
Their null geodesic equations along the radial direction on the equatorial plane of the black hole can be analogously
 realized as particle motion in the effective potential.
Section III  explores  the effects of the black hole charge on the circular trajectories   via this effective potential.
 In particular, we  solve the geodesic equations to
find  the radius of innermost circular trajectories and its corresponding impact parameter in terms  of  the black hole's parameters.
Section IV derives a closed-form expression for the equatorial light deflection angle. We have verified that, by taking the limit of $Q=0$,
our result reduces to the case of the Kerr black hole obtained by work \cite{SVI1}.
All results will be summarized in the closing section.

\section{Geodesic equations and innermost circular trajectories of light rays in Kerr-Newman spacetime}

 \begin{figure}[h]
 \centering
 \includegraphics[width=0.9\columnwidth=0.6]{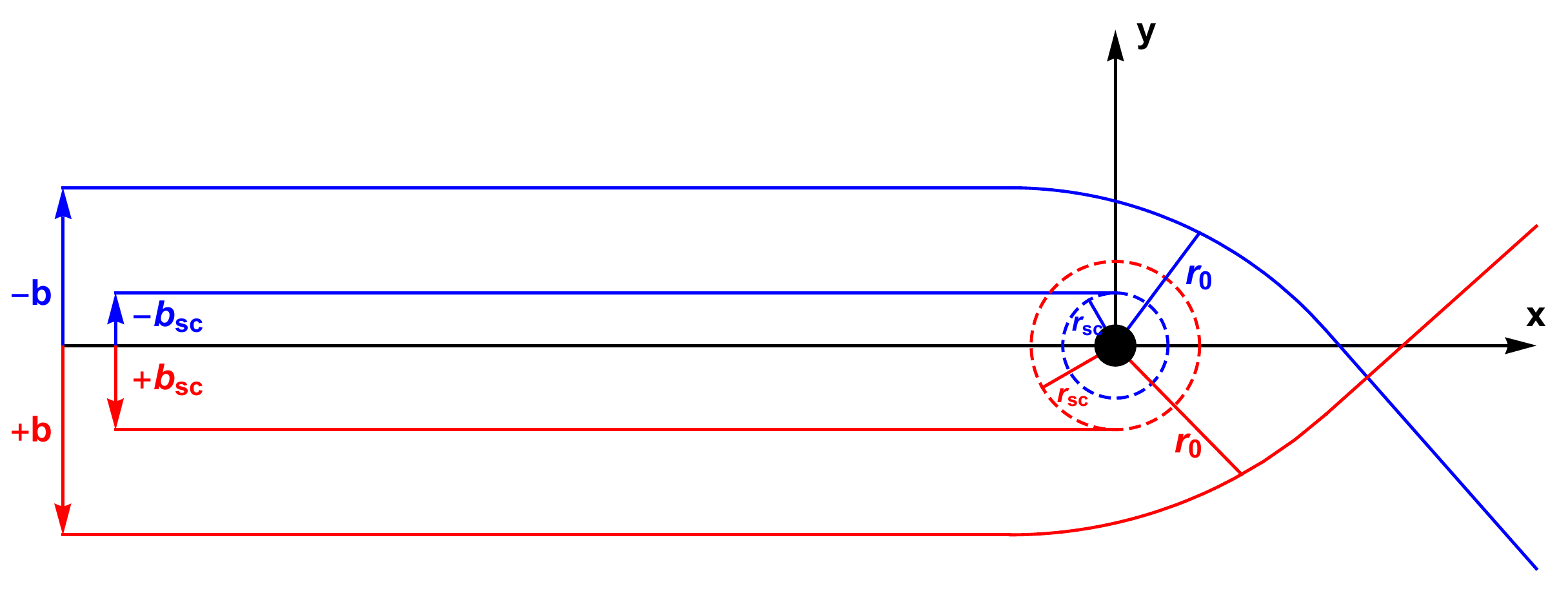}
 \caption{
 %\textcolor{blue}
 {Sign convention for orbits as viewed from above. The spin axis of the black hole points out of the page in this figure. The red (blue)
 solid line shows the direct (retrograde) orbit with the radius of closest approach $r_0$. The suffix "c" is added in the case of the
 innermost circular motion.} } \label{fig:int_KN}
 \end{figure}

In this paper, we thoroughly study the light bending due to the Kerr-Newman black hole, in which spacetime outside a black hole with the
gravitational mass $M$, charge $Q$, and  angular momentum per unit mass $a=J/M$ is described
 by the Kerr-Newman metric as
\ba
 {ds}^2 &=& g_{\mu\nu} dx^\mu dx^\nu  \nonumber \\
&= & -\frac{ \left(\Delta -a^2 \sin^2\theta \right)}{\Sigma } {dt}^2   + \frac{ a   \sin ^2\theta \left(Q^2-2 M
   r\right)}{\Sigma } ({dt}{d\phi+ d\phi dt)}  \nonumber\\
&&\quad\quad+  \frac{\Sigma}{\Delta} dr^2 +\Sigma{\, d\theta}^2 +\frac{ \sin ^2 \theta}{\Sigma} \left(( r^2+a^2)^2 -a^2 \Delta  \sin
   ^2\theta \right) {d\phi}^2 \, , \label{KN_metric}
   \ea
where
\be
\Sigma=r^2+a^2\cos^2\theta \, , \quad  \Delta=r^2+a^2+Q^2-2 M r\,.
\ee
The  event horizon $R_{H}$ can be found by solving $\Delta(r)=0$, and is given by
\be
R_{H}=M+\sqrt{M^2-(Q^2+a^2)}\,
\ee
with the condition $M^2 > Q^2+ a^2$. The Lagrangian  of a particle is then
 \begin{align}
 \mathcal{L}=\frac{1}{2}g_{\mu\nu}u^\mu u^\nu \,
 \end{align}
with the 4-velocity  $u^\mu=dx^\mu/d\lambda$ defined in terms of   an affine parameter $\lambda$.

Due to the fact that the metric of the Kerr-Newman black hole is independent of $t$ and $\phi$, the associated Killing vectors are
 $\xi_{(t)}^\mu$ and $\xi_{(\phi)}^\mu$ given, respectively, by
 \begin{align}
 \xi_{(t)}^\mu=\delta_t^\mu , \quad \xi_{\phi}^\mu=\delta_\phi^\mu \,.
 \end{align}
Then, together with the 4-velocity of  light rays,
the  conserved quantities, namely energy and azimuthal angular momentum, along a geodesic, can be constructed by the above Killing
vectors
 \begin{align}
 \varepsilon & \equiv -\xi_{(t)}^\mu u_\mu=\frac{1}{\Sigma}\left[a\left(\ell-\varepsilon
 a\sin^2{\theta}\right)+\frac{(r^2+a^2)\left[\varepsilon(r^2+a^2)-a\ell\right]}{\Delta}\right] \; ,\\
\ell & \equiv \xi_{\phi}^\mu u_\mu=\frac{1}{\Sigma}\left[\frac{\ell-a\varepsilon\sin^2{\theta}}{\sin^2{\theta}}
+\frac{a\left[\varepsilon(r^2+a^2)-a\ell\right]}{\Delta}\right] \; ,
 \end{align}
where $\varepsilon$ and $\ell$ are the light ray's energy and azimuthal angular momentum evaluated at spatial infinity. Light rays
travel along null world lines obeying the condition
 $ u^\mu u_\mu=0$.
Additionally, there exists a Carter constant
\begin{align}
 \kappa = u_\mu u_\nu K^{\mu \nu} -(\ell- a \varepsilon)^2 \, ,
\end{align}
where
 \begin{eqnarray}
 K^{\mu \nu}&=& \Delta \, k^{ \mu } q^{\nu } + r^2 g^{\mu \nu} \, , \nonumber\\
 q^{\mu} &=&= \frac{1}{\Delta} [(r^2+a^2) \delta^{\mu}_t+\Delta \delta^{\mu}_r + a \delta^{\mu}_\phi ] \, ,\nonumber\\
 k^{\mu} &=&= \frac{1}{\Delta} [(r^2+a^2) \delta^{\mu}_t-\Delta \delta^{\mu}_r + a \delta^{\mu}_\phi ] \, .
 \end{eqnarray}
 Using these three constants of motion together with  $u_{\mu} u^{\mu}=0${, we are able to write down the general geodesic equations for
 light rays  in terms of $\varepsilon$, $\ell$, and $\kappa$ as
 \begin{align}
 \Sigma\dot{t} & =-a\left(a\varepsilon\sin^2{\theta}-\ell\right)+\frac{(r^2+a^2)
 \left[\varepsilon(r^2+a^2)-a\ell \right]}{\Delta} \, , \\
 \Sigma\dot{\phi} & =-\left(a\varepsilon-\frac{\ell}{\sin^2{\theta}}\right)
 +\frac{a \left[\varepsilon(r^2+a^2)-a\ell \right]}{\Delta} \, , \label{phi_dot}\\
 \Sigma^2\dot{r}^2 & =(\varepsilon(r^2+a^2)-a\ell )^2-\Delta\left[(\ell-a\varepsilon)^2+\mathcal{\kappa}\right] \, , \label{r_dot}\\
 \Sigma^2\dot{\theta}^2 & =\kappa+\cos^2{\theta}\left(a^2\varepsilon^2-\frac{\ell^2}{\sin^2{\theta}}\right) \label{thetadot}\,.
 \end{align}
The overdot means the derivative with respect to the affine parameter $\lambda$.
To indicate whether the light ray is traversing along the direction of frame dragging or opposite to it,
we define the following impact parameter :
 \begin{align}
 b_s=s\left|\frac{\ell}{\varepsilon}\right|\equiv s\,b \; ,
 \end{align}
where $s=\text{Sign$(\ell/\varepsilon)$}$ and $b$ is the positive magnitude. The parameter $s=+1$ for $b_s>0$ will be referred to as
direct orbits; and those with $s=-1$ for $b_s<0$ as retrograde orbits (see Fig.\ref{fig:int_KN} for the sign convention).

Here we restrict the light rays traveling on the equatorial plane of the black hole
 by choosing $\theta={\pi}/{2}$, and $\dot{\theta}=0$, so that $\kappa=0$ in Eq.~(\ref{thetadot}).
Rewriting the equation of motion along the radial direction, \eqref{r_dot}  allows us to define the function $W_\text{eff}$ from
 \begin{align}
 \frac{1}{b^2} & =\frac{\dot{r}^2}{\ell^2}+W_\text{eff}(r)\; ,
 \end{align}
where
 \begin{align}
 W_\text{eff}(r)=\frac{1}{r^2}\left[1-\frac{a^2}{b^2}+\left(-\frac{2M}{r}+\frac{Q^2}{r^2}\right)\left(1-\frac{a}{b_s}\right)^2\right] \,
 . \label{eq:Weff}
 \end{align}
 The above equation is analogous to that of particle motion in the effective potential $ W_\text{eff}(r)$ shown in Fig.\ref{fig:Weff}
 with the kinetic energy ${\dot r}^2/\ell^2$ and constant total energy $1/b^2$ ~\cite{HAR,VRI4}.
 \begin{figure}[h]
 \centering
 \includegraphics[width=0.8\columnwidth=0.5]{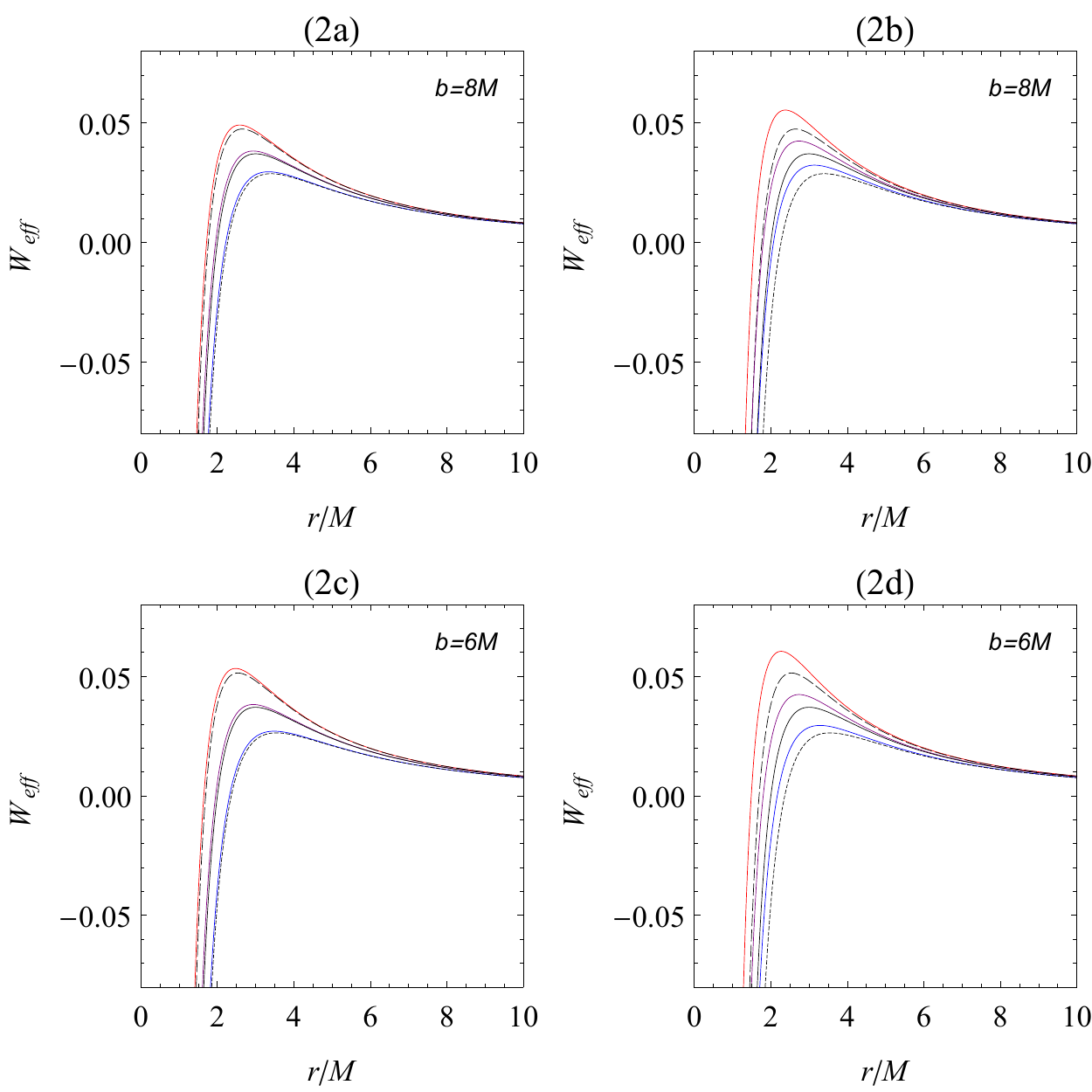}
 \caption{
 %\textcolor{blue}
 {
 The the ``effective potential" $W_{\rm eff}$ as a function of $r/M$ for four sets of parameters.
(a) $a/M=0.5$, $Q/M=0.3$, and $b=8M$; (b) $a/M=0.5$, $Q/M=0.6$, and $b=8M$; (c) $a/M=0.5$, $Q/M=0.3$, and $b=6M$; (d) $a/M=0.5$,
$Q/M=0.6$, and $b=6M$.
 The plot convention used henceforth: Kerr-Newman direct (red), Kerr-Newman retrograde (blue), Kerr direct (dashed with $Q=0$), Kerr
 retrograde (dotted with $Q=0$), Reissner-Nordstrom (purple with $a=0$), and Schwartzschild (black with $Q=0, a=0$).
 %\textcolor{blue}
 {The nonzero charge of the black hole  gives more repulsive effects  to the light rays as seen from the effective
 potential than the corresponding  black hole with zero $Q$. }
 %where the other parameters are chosen as above.
 %(2a) Kerr-Newman direct with $a/M=0.5$ and $Q=M=0.3$ (red), Kerr-Newman retrograde with $a/M=0.5$ and %$Q=M=0.3$ (blue), Kerr direct with $a/M=0.5$ (dashed), Kerr retrograde with $a/M=0.5$ and $Q=M=0.3$ (dotted), %Reissner-Nordstrom with $Q/M=0.3$ (purple), andSchwartzschild (black). The impact parameter is $b=8M$
 }
 \label{fig:Weff} }
 \end{figure}

 In the limits of $Q=0$ and $Q=0,\;a=0$, the effective potential $W_{\rm eff}$ reduces to the {ones for the} Kerr and Schwarzschild
 black holes respectively~\cite{HAR}.
In the case of the  Kerr-Newman black hole,
 the nonzero charge of the black hole seems to give repulsive effects  to the light rays as seen from its contributions to the
 centrifugal potential of the $1/r^2$ term. These repulsive forces in turn affect light rays so as to prevent them from collapsing into
 the event horizon, and thus
 as will be seen later, shift  the innermost
  circular trajectories  of the light rays  toward the black hole.
Also,  the presence of the black hole's charge is found to decrease the deflection angle due to this additional repulsive effect on the
light rays, as  compared with Kerr's case for  the same impact parameter $b$.

To see it, let us consider a light ray that starts in the asymptotic region to approach the black hole, and then turn back to the
asymptotic region to reach the observer. Such light rays have a turning point, the radius of closest approach to a black hole $r_0$,
which  crucially depends on the impact parameter $b$, determined by
 \begin{align}
 \left.\frac{\dot{r}^2}{\ell^2}\right|_{r=r_0}=\frac{1}{b^2}-W_\text{eff}(r_0)=0 \,. \label{eq:rdot_Weff}
 \end{align}
Equation~\eqref{eq:rdot_Weff} leads to a quartic equation in $r_0$
 \begin{align}
 r_0^4-b^2\left(1-\frac{a^2}{b^2}\right)r_0^2+2Mb^2\left(1-\frac{a}{b_s}\right)^2 r_0-Q^2b^2\left(1-\frac{a}{b_s}\right)^2=0 \,
 .\label{eq:TDOCA}
 \end{align}
We express the solutions in terms of trigonometric functions, where one of the three roots of~\eqref{eq:TDOCA} gives the analytical
result of $r_0$ as
\begin{align}
 r_0(b_s)=\frac{b}{\sqrt{6}}\sqrt{1-\omega_s^2} \Bigg   \{ & \sqrt{1+\sqrt{1-\Omega_s}\cos{\left(\frac{2\Theta}{3}\right)}} \no\\
 &+\sqrt{2-\sqrt{1-\Omega_s}\cos{\left(\frac{2\Theta}{3}\right)}-\frac{3\sqrt{6}M(1-\omega_s)^2}{b(1-\omega_s^2)^{\frac{3}{2}}\sqrt{1+\sqrt{1-\Omega_s}\cos\left(\frac{2\Theta}{3}\right)}}}\Bigg\}
 \,  \label{eq:r0}
 \end{align}
with $\omega_s$,\, $\Omega_s$ and $\Theta$, depending explicitly on the black hole's parameters,
as well as the impact parameter $b_s$, defined as
 \begin{align}
 \omega_s & =\frac{a}{b_s} \,,\no\\
 \Omega_s &=\frac{12\, Q^2}{b^2\, (1+\omega_s)^2} \, ,\no\\
 \Theta & =\arccos{\left(\frac{3\sqrt{3}M\, (1-\omega_s)^2}{b\, (1-\omega_s^2)^{\frac{3}{2}}\,
 (1-\Omega_s)^\frac{3}{4}}\sqrt{1-\frac{b^2\, (1+\omega_s)^3}{54M^2\,
 (1-\omega_s)}\left[1+3\,\Omega_s-(1-\Omega_s)^{\frac{3}{2}}\right]}\right)}
\,. \end{align}
{In the case of the Kerr black hole with $Q=0$, we have}
{\begin{align*}
 \Omega_s & =0 \, ,\\
 \Theta & =\arccos{\left(\frac{3\sqrt{3}M(1-\omega_s)^2}{b(1-\omega_s^2)^{\frac{3}{2}}}\right)} \, .
 \end{align*}}
Equation {\eqref{eq:r0} reduces to $r_0(b)$ in~\cite{SVI1}} as
\begin{align*}
 r_0(b_s) &
 =\frac{b}{\sqrt{6}}\sqrt{1-\omega_s^2}\left\{\sqrt{1+\cos{\left(\frac{2\Theta}{3}\right)}}+\sqrt{2-\cos{\left(\frac{2\Theta}{3}\right)}-\frac{3\sqrt{6}M(1-\omega_s)^2}{b(1-\omega_s^2)^{\frac{3}{2}}\sqrt{1+\cos{\left(\frac{2\Theta}{3}\right)}}}}\right\}
 \\
 &
 =\frac{2b}{\sqrt{3}}\sqrt{1-\frac{a^2}{b^2}}\cos{\left[\frac{1}{3}\arccos{\left(\frac{-3\sqrt{3}M}{b}\frac{\left(1-\frac{a}{b_s}\right)^2}{\left(1-\frac{a^2}{b^2}\right)^{\frac{3}{2}}}\right)}\right]}
\, .
\end{align*}
The distance of closest approach $r_0$ for the light ray to travel around Kerr-Newman black holes, given in Eq.\eqref{eq:r0}, certainly
generalizes that of the Kerr or the Schwarzschild black holes depicted in Fig.\ref{fig:r0}.

 \begin{figure}[h]
 \centering
 \includegraphics[width=0.9\columnwidth=0.90]{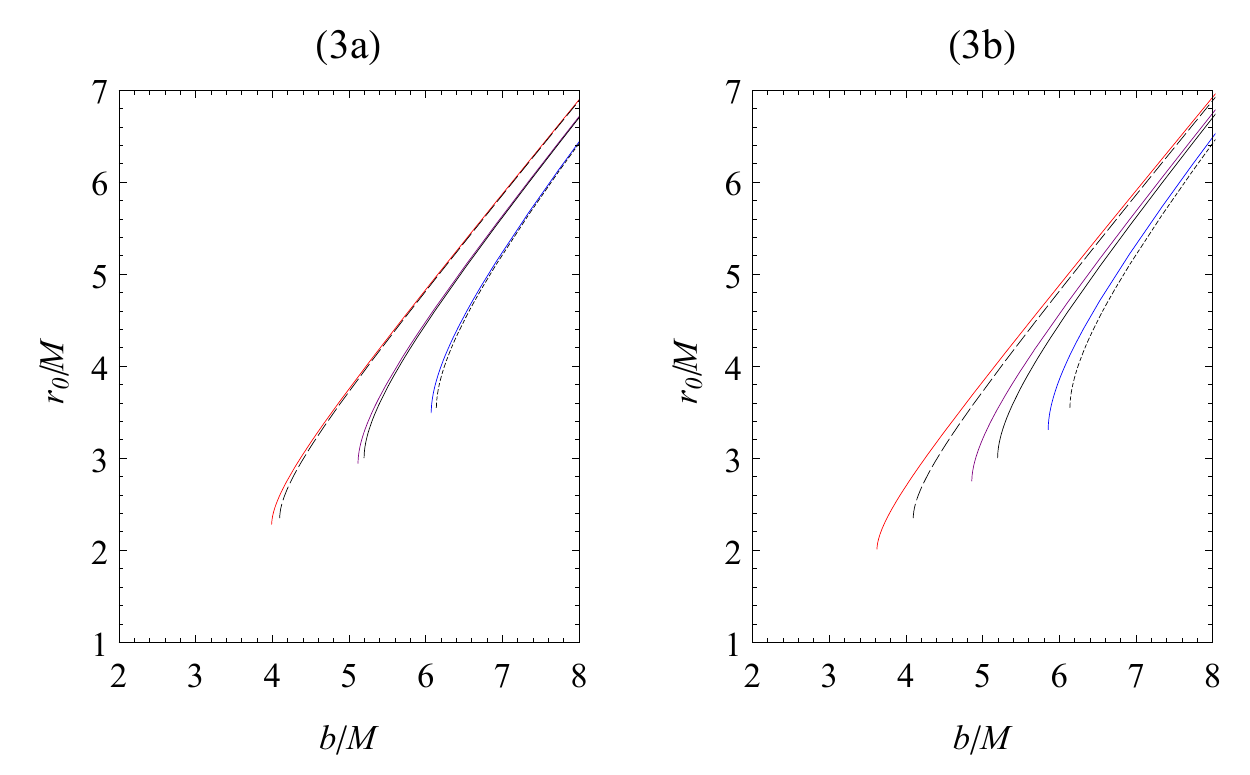}
 \caption{
 %\textcolor{blue}
{ The distance of closest approach $r_0/M$ as a function of the impact parameter $b/M$ for (a) $a/M=0.5$ and $Q/M=0.3$; (b) $a/M=0.5$
and $Q/M=0.6$. The plots show the results for the Schwarzschild, Reissner-Nordstr\"om, Kerr and Kerr-Newman black holes for comparison.
%\textcolor{blue}
{The additional repulsive effects from the charge of the black hole pushes the distance of closest approach $r_0$ of
light rays being away from the black hole for a fixed  impact parameter $b$.} } } \label{fig:r0}
 \end{figure}
 It is anticipated that for both direct and retrograde motions, the repulsive effect from the charge of black holes pushes the distance
 of closest approach $r_0$ being away from the black hole for a fixed  impact parameter $b$, which can be compared with the Kerr case
 with the same impact parameter $b$.

\section{Critical impact parameters and innermost circular orbits}

Again, consider the light rays coming in from spatial infinity with the impact parameter $b$.
The plots in Fig. \ref{fig:Weff} show the shape of the effective potential $W_\text{eff}(r)$ that vanishes at large $r$ and has one
maximum.
The behavior of the light ray trajectories depends on whether $1/b^2$ is greater or less than the maximum height of $W_\text{eff}(r)$.
The innermost trajectories of light rays  have a direct consequence on  the apparent shape of the black hole with
the smallest radius ${r_c}$ when the turning point $r_0$ is located at the maximum of $W_\text{eff}(r)$ for a particular choice of
${b_{c}}$  obeying
 \begin{align}
 \left.\frac{d\,W_\text{eff}(r)}{dr}\right|_{r=r_{sc} } & =0  \;
 \end{align}
with the value
 \begin{align}
r_{sc}=\frac{3M}{2}\left(\frac{1-\frac{a}{b_{sc}}}{1+\frac{a}{b_{sc}}}\right)\left[1+\sqrt{1-\frac{8Q^2\left(1+\frac{a}{b_{sc}}\right)}{9M^2\left(1-\frac{a}{b_{sc}}\right)}}\right]
\, ,\label{fun:rc}
 \end{align}
 which is a circular motion forming a photon sphere.
 However, these circular trajectories are unstable because any small change in $b$ results in the trajectory moving away from the
 maximum.
 The radius of the circular photon orbit ${r_{sc}}$ above is
 consistent with the finding in \cite{CHAS}, and
 the known result of  $r_{sc}$ in the limit $Q=0$ given in~\cite{SVI1}.
 We then substitute \eqref{fun:rc} into \eqref{eq:TDOCA} for obtaining the corresponding critical impact parameter
 ${b_{sc} }$.

 Substituting (\ref{fun:rc}) into (\ref{eq:TDOCA}), it is  more convenient to express the equations in terms of $y_+$ and $y_-$, respectively \cite{SVI1},
 \begin{align}
 y_+ & =b_{+c}+a \, , \label{yb_a+} \\
 y_- & =-(b_{-c}+a) \, ,\label{yb_a}
 \end{align}
 which obey the quartic equations
\begin{align}
 & (Q^2-M^2)y_+^{4}+2M^2ay_+^{3}+(27M^4-36M^2Q^2+8Q^4)y_+^{2} \no\\
 & \quad\quad\quad\quad\quad - (108M^4a-72M^2Q^2a)y_{+}+(108M^4a^2+16Q^6)=0 \, , \label{eq:bsc-gen-p}\\
 & (Q^2-M^2)y_-^{4}-2M^2ay_-^{3}+(27M^4-36M^2Q^2+8Q^4)y_-^{2} \no\\
 & \quad\quad\quad\quad\quad + (108M^4a-72M^2Q^2a)y_{-}+(108M^4a^2+16Q^6)=0 \, . \label{eq:bsc-gen-m}
\end{align}
 The  solutions of the critical value ${b_{sc} }$  in a Kerr-Newman black hole are found analytically to be
  \begin{align} \label{bsc}
 b_{sc} & =-a+\frac{M^2a}{2(M^2-Q^2)}+\frac{s}{2\sqrt{3}(M^2-Q^2)}\Bigg[\sqrt{V+(M^2-Q^2)\left(U+\frac{P}{U}\right)} \no\\
 &
 +\sqrt{2V-(M^2-Q^2)\left(U+\frac{P}{U}\right)-\frac{s6\sqrt{3}M^2a\left[(M^2-Q^2)(9M^2-8Q^2)^2-M^4a^2\right]}{\sqrt{V+(M^2-Q^2)\left(U+\frac{P}{U}\right)}}}\Bigg]
 \end{align}
where
 \begin{align}
 P & =(3M^2-4Q^2)\left[9(3M^2-4Q^2)^3+8Q^2(9M^2-8Q^2)^2-216M^4a^2\right] \, , \no\\
 U & =\bigg\{-\left[3(3M^2-2Q^2)^2-4Q^4\right]\left[9M^2(9M^2-8Q^2)^3-8\left[3(3M^2-2Q^2)^2-4Q^4\right]^2\right] \, \no\\
 &\qquad +108M^4a^2\left[9(3M^2-4Q^2)^3+4Q^2(9M^2-8Q^2)^2-54M^4a^2\right]\,  \no\\
 &\qquad +24\sqrt{3}M^2\sqrt{(M^2-a^2-Q^2)\left[Q^2(9M^2-8Q^2)^2-27M^4a^2\right]^3}\bigg\}^\frac{1}{3} \, ,\no\\
 V & =3M^4a^2+(M^2-Q^2)\left[6(3M^2-2Q^2)^2-8Q^4\right] \,.
 \end{align}

 \begin{figure}[h]
 \centering
 \includegraphics[width=0.9\columnwidth=0.9]{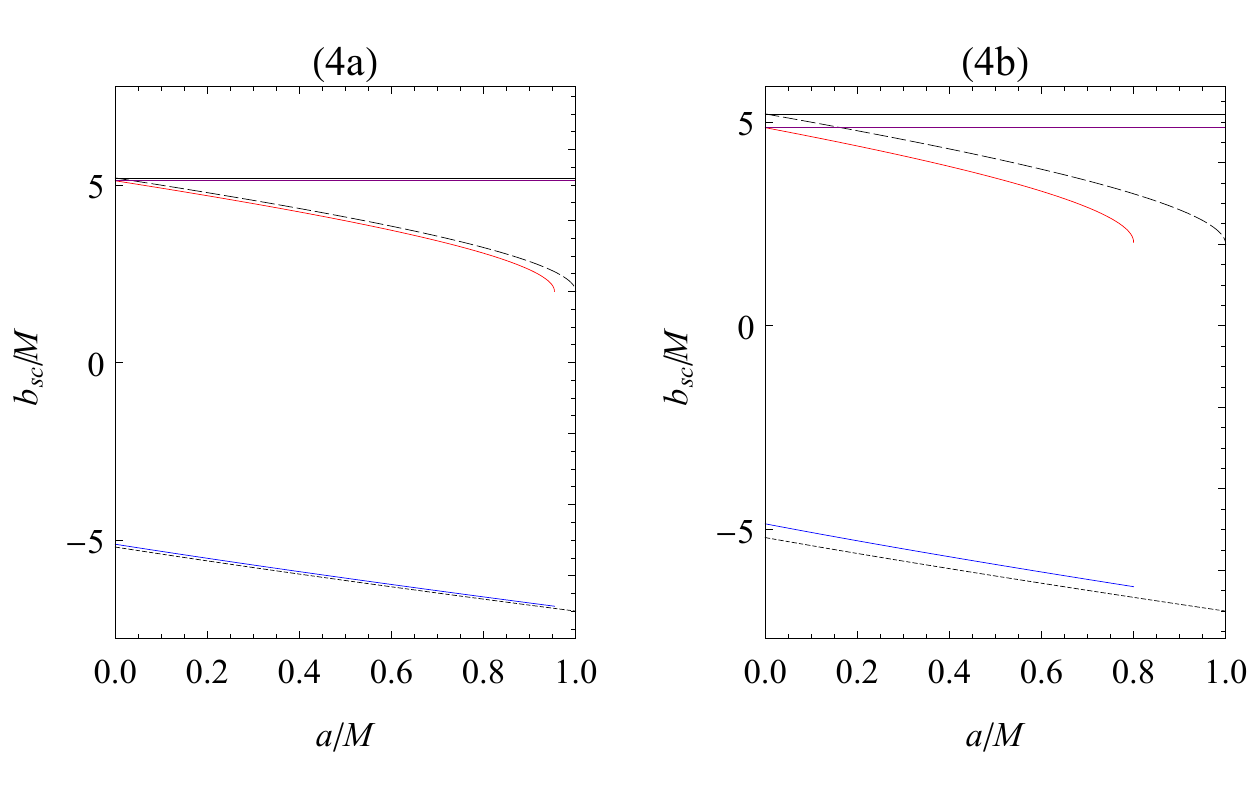}
 \caption{
 %\textcolor{blue}
 {The critical impact parameter $b_{sc}/M$ as a function of the spin parameter $a/M$ for (a) $Q/M=0.3$, (b) $Q/M=0.6$.
The plots show the Schwarzschild, Reissner-Nordstrom, Kerr and Kerr-Newman black holes for comparison.
%\textcolor{blue}
{The circular
orbits exist for a charged black hole  with the smaller impact parameter $\vert b_{sc} \vert$ as compared with the corresponding  black
hole with $Q=0$ for the same $a$.} }} \label{fig:bsca}
 \end{figure}
 \begin{figure}[h]
 \centering
 \includegraphics[width=0.9\columnwidth=0.9]{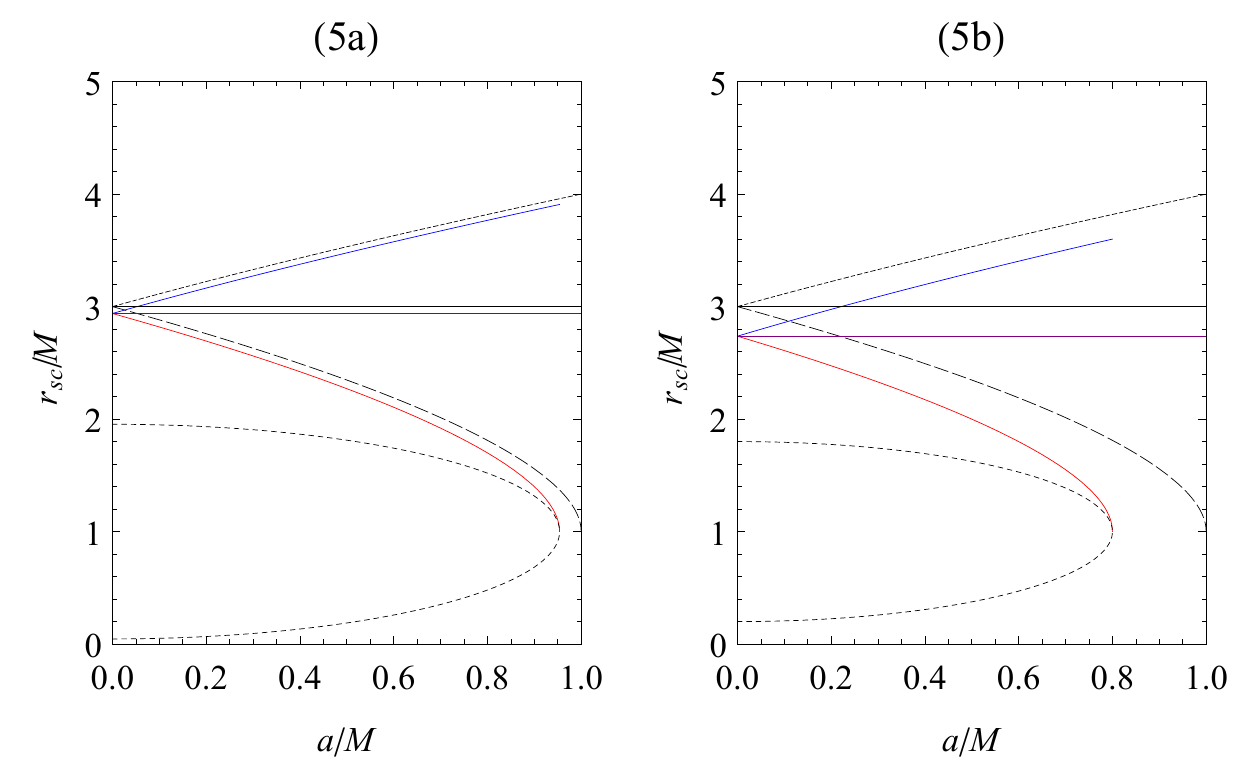}
 \caption{
 %\textcolor{blue}
 {
 The critical radius $r_{sc}/M$  plotted as a function of the spin parameter for (a) $Q/M=0.3$, (b) $Q/M=0.6$.
 The cases of  $r_{sc}$ of the Schwarzschild, Reissner-Nordstrom, Kerr and Kerr-Newman black holes are drawn for comparison.
 Additionally, the Kerr-Newman outer(inner) horizon radii are shown in red(orange) dashed line for reference.
 %\textcolor{blue}
 {The
 circular orbits exist for a charged black hole with a smaller value of the radius $r_{sc}$ as compared with the corresponding  black
 hole with $Q=0$ for a fixed $a$.}} }\label{fig:rsca}
 \end{figure}

 %========================================================================================================%
 \begin{figure}[h]
 \centering
 \includegraphics[width=0.9\columnwidth=0.9]{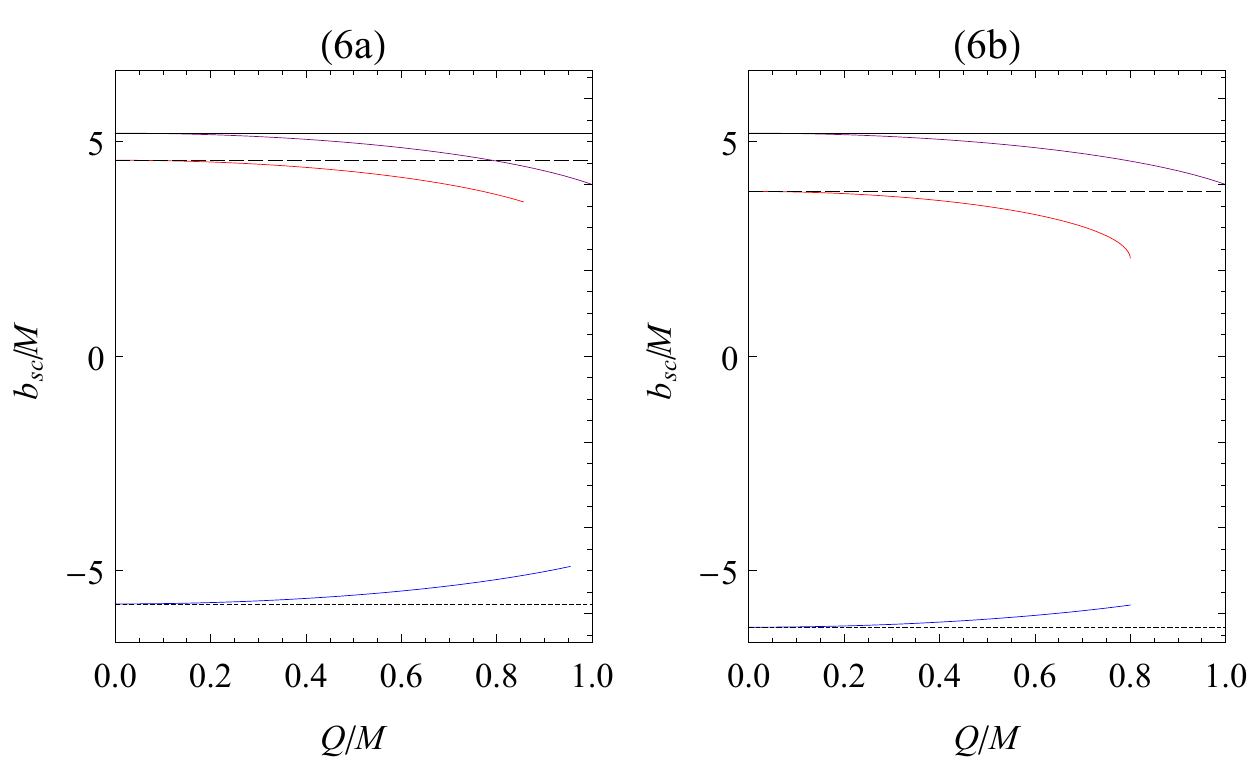}
 \caption{
%\textcolor{blue}
{ The critical impact parameter $b_{sc}/M$ as a function of charge $Q/M$ for (a) ${a/M}=0.3$, (b) ${a/M}=0.6$. The plots show the
results for the Schwarzschild, Reissner-Nordstr\"om, Kerr and Kerr-Newman black holes for comparison.
%\textcolor{blue}
{The critical
impact parameter $\vert b_{sc}\vert$ decreases as charge $Q$ of the black hole increases.}}} \label{fig:bscQ}
 \end{figure}
 \begin{figure}[h]
 \centering
 \includegraphics[width=0.9\columnwidth=0.9]{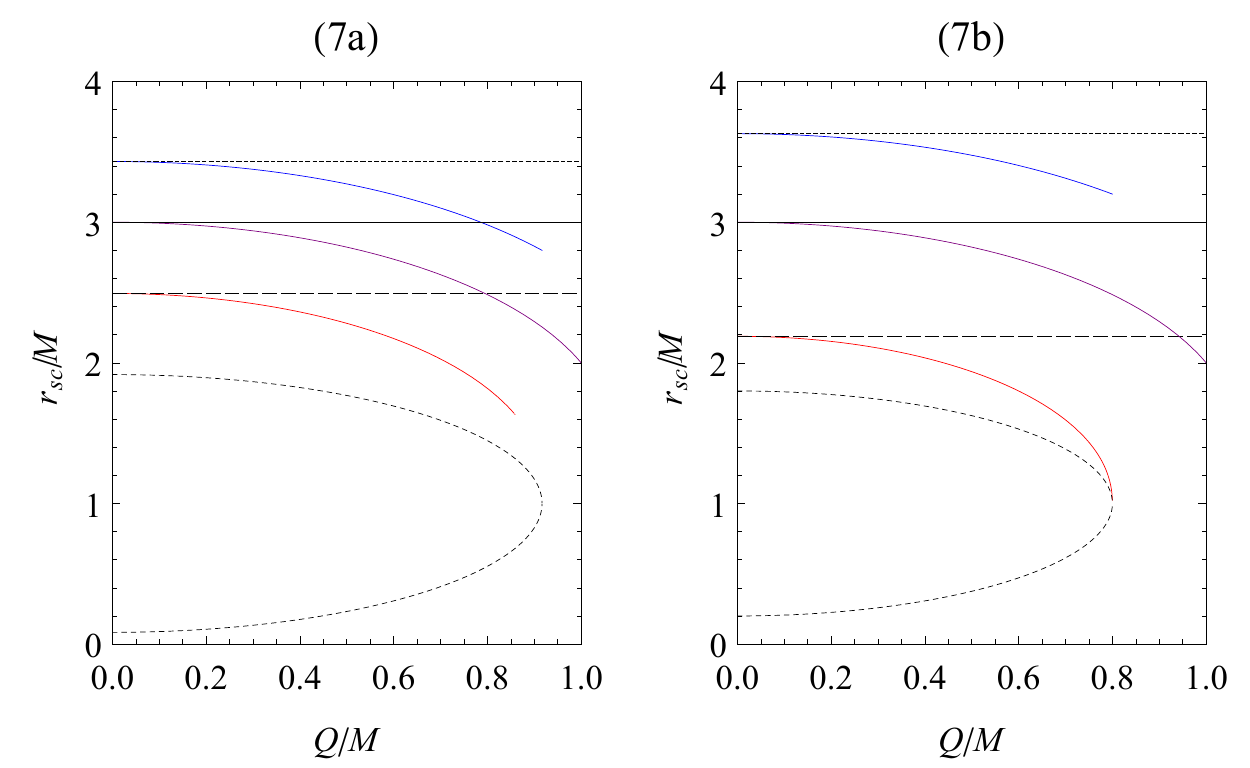}
 \caption{
 %\textcolor{blue}
 {The critical radius $r_{sc}/M$ as a function of charge ${Q/M}$ for (a) ${a/M}=0.3$, (b) ${a/M}=0.6$.
 The plots of the Schwarzschild, Reissner-Nordstr\"om, Kerr and Kerr-Newman black holes are drawn for comparison.
 Additionally, the Kerr-Newman outer(inner) horizon radii are also shown in red(orange) dashed line for reference.
 %\textcolor{blue}
 {The
 radius of the innermost circular motion of light rays $r_{sc}$  decreases as charge $Q$ of the black hole increases.}}}
 \label{fig:rscQ}
 \end{figure}
 %%========================================================================================================%

Although  one can  numerically check the consistency between  the expression of (\ref{bsc}) in the limit of $Q=0$ and that of the Kerr
case given by \cite{SVI1}
\begin{align} \label{bsc_kerr}
 b_{sc} & \rightarrow -a+ s6 M \cos\bigg[\frac{1}{3} \cos^{-1} \bigg( \frac{-sa}{M} \bigg) \bigg] \, ,
 \end{align}
direct simplification to recover (\ref{bsc_kerr}), by setting $Q=0$ in (\ref{bsc}), is not so trivial to achieve.
An alternative consistency check is to consider that
 {the above two quartic equations (\ref{eq:bsc-gen-p}) and (\ref{eq:bsc-gen-m}) in the limit of Kerr case, $Q=0$  lead to
{\begin{align}
 & -M^2\left(y_+-2a\right)\left(y_+^3-27M^2y_++54M^2a\right)=0  \, ,\\
 & -M^2\left(y_-+2a\right)\left(y_-^3-27M^2y_--54M^2a\right)=0 \, .
 \end{align}}
Their solutions
 certainly  give (\ref{bsc_kerr}) using (\ref{yb_a+}) and (\ref{yb_a}). Then, the radius of innermost circular trajectories ${r_{sc} }$ with the impact
 parameter ${b_{sc} }$ can be obtained  through (\ref{fun:rc}), which will be a tedious function of the black hole's parameters.

According to \cite{CHAR},  in fact, the equation to determine the radius of innermost circular motions
${r_{c}}$ in terms of the black hole's parameters can be derived directly.
Let us consider a particle with mass $m$ moving around the Kerr-Newman black hole.
There exists the circular motion of the particle with the radius $r$ when the energy $E$ and azimuthal angular momentum $L$
satisfy \cite{DAD,LIU}
\ba
\frac{E}{m} = \frac{ a \sqrt{ M r-Q^2} + (Q^2+r^2-2 M r)}{ r \sqrt{ 2 Q^2 + r^2- 3 M r + 2 a ( M r-Q^2)^{1/2}}} \, ,\label{E} \\
\frac{L}{m}=\frac{ a (Q^2 -2 M r)+ ( a^2 + r^2) \sqrt{M r -Q^2}}{ r \sqrt{ 2 Q^2 + r^2- 3 M r + 2 a ( M r-Q^2)^{1/2}}} \, .\label{L}
\ea
%
%The radius of the last circular orbit $r_{lc}$ is mainly determined when both $\varepsilon_o$ and $\ell_o$ become infinity at $r_o=
%r_{lc}$ given by
%\be
%2 a (M r_{lc}-Q^2)^{1/2}+2Q^2+r_{lc}(r_{lc}-3M) = 0 \,
%\ee
%\textcolor{red}{The typical behavior of $R (r_o)$ as a function of $r_o$ is plotted in Fig.(\ref{R_ro}).}
%\begin{figure}
 % \centering
 % \includegraphics[scale=0.45]{iscovsavq2.png}
 % \caption{\textcolor{red}{The behavior of $R(r_o)$ in~(\ref{R}) with the input  of $\ell_o$~(\ref{l_o}) and $\epsilon_o$~(\ref{e_o})is shown for the black hole of $Q=0.2 M$ and $a=0. 96 M$. }} \label{R_ro}
%\end{figure}
%
%
Apparently, the radius of the circular motion cannot be arbitrarily small.
In particular, in the case of massless limit $( m \rightarrow 0)$, the conditions of having finite values of $E$ and $L$ require  the
radius of the circular photon orbit obeying
\begin{align}
 2 Q^2 + r_c^2- 3 M r_c + 2 a ( M r_c-Q^2)^{1/2}=0 \, .
\end{align}
 In the limit of $Q=0$, we have one of the roots given by
\begin{align}
r_{sc}=  2 M\bigg\{ 1+ \cos\bigg[\frac{2}{3} \cos^{-1} \bigg( \frac{-sa}{M} \bigg) \bigg]\bigg\}
\;
\end{align}
in agreement  with the Kerr case in \cite{CHAS,SVI1}.
In the general situation  of finite $Q$, the solution of $r_c$ becomes
\begin{align}
 r_{sc} & =\frac{3M}{2}+\frac{1}{2\sqrt{3}}\sqrt{9M^2-8Q^2+U_{c}+\frac{P_{c}}{U_{c}}} \no\\
 &\quad -\frac{s}{2}\sqrt{6M^2-\frac{16Q^2}{3}-\frac{1}{3}\left(U_c+\frac{P_{c}}{U_{c}}\right)
 +\frac{8\sqrt{3}Ma^2}{\sqrt{9M^2-8Q^2+U_c+\frac{P_c}{U_c}}}} \;\; ,
 \label{rc}
 \end{align}
where
 \begin{align}
 P_{c} & =(9M^2-8Q^2)^2-24a^2(3M^2-2Q^2) \, , \no\\
 U_{c} & =\bigg\{(9M^2-8Q^2)^3-36a^2(9M^2-8Q^2)(3M^2-2Q^2)+216M^2a^4 \no\\
 &\quad\quad +24\sqrt{3}a^2\sqrt{(M^2-a^2-Q^2)\left[Q^2(9M^2-8Q^2)^2-27M^4a^2\right]}\bigg\}^\frac{1}{3} \, .
 \end{align}
%

%\textcolor{blue}
{For the Reissner-Nordstrom black holes,
 $a\to 0$, we find
 \begin{align*}
 P_{c} & = U_{c}^2=(9M^2-8Q^2)^2 \,
 \end{align*}
 with
 \begin{align} \label{r_cNR}
 r_{c} & =\frac{3M}{2}\left(1+\sqrt{1-\frac{8Q^2}{9M^2}}\right)
 \end{align}
as anticipated in \cite{CHAS}.}

Combining (\ref{fun:rc}) with (\ref{eq:TDOCA}) can derive the following useful relation
\begin{align}
 b_{sc} & =a+\frac{r_{sc}^2}{\sqrt{Mr_{sc}-Q^2}} \, .
 \end{align}
Thus, substituting the solution of $r_{sc}$ in (\ref{rc}) into the above relation can obtain
the result of $b_{sc}$ in terms of the black hole's parameters instead.

Plugging in the value for all parameters, we reproduce the critical impact parameter $b_{sc}$, and the corresponding radius of the innermost circular motion $r_{sc}$ in Figs.\ref{fig:bsca} and \ref{fig:rsca}, plotted as a function of $a$ for the Kerr black hole \cite{SVI1}, and also for the the Kerr-Newman black holes. Due to the fact that the charge of a black hole gives the repulsive effects to the light rays that prevent them from collapsing into the black hole, it is found that  the circular orbits exist for a smaller value of the radius $r_{sc}$ with the smaller impact parameter $ \vert b_{sc}\vert $ as compared with the Kerr case for the same $a$.
Also, the radius of the innermost circular motion of light rays with the critical impact parameter decreases as charge $Q$ of the black hole increases for both direct and retrograde motions seen in Figs.\ref{fig:bscQ} and \ref{fig:rscQ}.

\section{The exact expression of equatorial light deflection angle}

To obtain the closed-form deflection angle of a light ray due to the Kerr-Newman black hole,
we introduce
\begin{equation}
u=1/r
\end{equation}
and rewrite (\ref{phi_dot}) and (\ref{r_dot}), with further
combinations, as
 \begin{align}
 \left(\frac{du}{d\phi}\right)^2=\left[\frac{1-2Mu+(a^2+Q^2)u^2}
 {1-(2Mu-Q^2u^2)\left(1-\frac{a}{b_s}\right)}\right]^2 B(u) \label{eq:ODEdu/dphi}\;,
 \end{align}
where the quantity $B(u)$ is a quartic polynomial
 \begin{align}
 B(u)=-Q^2\left(1-\frac{a}{b_s}\right)^2u^4+2M\left(1-\frac{a}{b_s}\right)^2u^3-\left(1-\frac{a^2}{b^2}\right)u^2+\frac{1}{b^2}
 \label{eq:B(u)}\, .
 \end{align}
 The deflection angle $\hat\alpha$ as the light ray proceeds in from spatial infinity and back out again is just twice the deflection
 angle
 from the turning point $r=r_0$ to infinity.  Integrating the equation of motion for $\phi$ in~\eqref{eq:ODEdu/dphi} and subtracting
 $\pi$
from it lead to
\begin{align}
 \hat{\alpha} & =-\pi+2\int_{0}^{1/r_0} \frac{1-(2Mu-Q^2u^2)\left(1-\frac{a}{b_s}\right)}{1-2Mu+(a^2+Q^2)u^2}
 \frac{\mathrm{d}u}{\sqrt{B(u)}} \;. \label{eq:alpha}
\end{align}
With $B(u)$
that generally has three real positive roots $u_2,u_3,$ and $u_4$ and one real negative root $u_1$, we can rewrite~\eqref{eq:B(u)} as
 \begin{align}
 B(u)=-Q^2(1-\omega_s)^2\,(u-u_1)(u-u_2)(u-u_3)(u-u_4) \label{B_fun}\;
 \end{align}
%\textcolor{blue}
{with again $\omega_s=a/b_s$.}
By extending the approach of \cite{SVI1}, we parametrize  the four roots of $B(u)=0$ as follows:
 \begin{align}
 u_1 & =\frac{X-2M-Y}{4Mr_0} \, ,  \\
 u_2 & =\frac{1}{r_0} \, , \\
 u_3 & =\frac{X-2M+Y}{4Mr_0} \, , \\
 u_4 & =\frac{2M}{Q^2}-\frac{X}{2Mr_0} \, ,\label{u4}
 \end{align}
where we have introduced two functions $X$ and $Y$, to be determined later.
We assume that the functions $X$ and $Y$ in the limit of $Q=0$
smoothly reduce to the corresponding functions in a Kerr black hole in~\cite{SVI1}.
% as it will be seen.
If so, by taking the $Q=0$ limit,  the function $B(u)$ in a Kerr black hole in \cite{SVI1} can be recovered from (\ref{B_fun}),
where the root $u_4$ is removed.
Comparing the coefficients of different powers of $u$ in $B(u)$ to those in the original polynomial in~\eqref{eq:B(u)}, we obtain the
equations to determine the functions $X$ and $Y$ as follows:
\begin{align}
 & Q^2\left[Y^2-(X-2M)(X+6M)+4X^2\right]=
 16M^2r_0 \left( X-r_0\frac{1+\omega_s}{1-\omega_s}\right) \, , \label{upow2} \\
 &Y^2-(X-2M)^2=\frac{8M(X-2M)(Q^2X-4M^2r_0)}{Q^2(X-2M)-4M^2r_0} \, , \label{upow1} \\
 & \left[ Y^2-(X-2M)^2\right] \left(\frac{1}{8Mr_0^3}-\frac{Q^2X}{32M^3r_0^4}\right)=\frac{1}{b^2(1-\omega_s)^2}\, . \label{upow0}
 \end{align}
Notice that these equations are given
in terms of the black hole's parameters as well as the distance of closest approach $r_0$ of a light ray with the impact parameter $b$.
 Substituting~\eqref{upow2} into~\eqref{upow0} reproduces  (\ref{eq:TDOCA}), which  shows that  $1/r_0$ is one of the roots.
 The combination of  \eqref{upow2} and \eqref{upow1}, with further rearrangement, gives a cubic equation of $X$,
 \begin{align}
 & \frac{Q^2}{2M}X^3-(Q^2+4Mr_0)X^2+\left(4M^2r_0+2MQ^2+\frac{8M^3r_0^{2}}{Q^2}+\frac{2Mr_0^{2}(1+\omega_s)}{(1-\omega_s)}\right)X \no\\
 & \quad\quad \quad\quad \quad\quad \quad
 \quad\quad=4M^2Q^2+\frac{4M^2r_0^{2}(1+\omega_s)}{1-\omega_s}+\frac{8M^3r_0^{3}(1+\omega_s)}{Q^2(1-\omega_s)} \, . \label{eq:X}
 \end{align}
We can solve directly the cubic equation~\eqref{eq:X}, and one of three roots is
 \begin{align}
  X(M,Q,r_0,\omega_s) &= \frac{2M(Q^2+4Mr_0)}{3Q^2}
 \no \\
 &+\frac{8M^2r_0}{3Q^2}\sqrt{1+\frac{Q^2}{2M^2r_0}\left(M-\frac{3r_0(1+\omega_s)}{2(1-\omega_s)}
 -\frac{Q^2}{r_0}\right)}\cos{\left(\frac{\Theta_X}{3}+\frac{2\pi}{3}\right)}
 \; , \label{eq:Xsol}
 \end{align}
where
\begin{equation}
 \Theta_X=\arccos{\left[\frac{-8M^3r_0^3-3MQ^2r_0^2\left(2M-\frac{3r_0(1+\omega_s)}
 {(1-\omega_s)}\right)-3Q^4r_0\left(5M-\frac{3r_0(1+\omega_s)}{(1-\omega_s)}\right)+10Q^6}
 {\left[4M^2r_0^2+Q^2r_0\left(2M-\frac{3r_0(1+\omega_s)}
 {(1-\omega_s)}\right)-2Q^4\right]^{\frac{3}{2}}}\right]}
\, .
\end{equation}
 The limit of $Q=0$ leads to $\Theta_X=\pi$. In this case Eq. (\ref{eq:Xsol}) can be simplified enormously as
 \begin{align}
 X(M,Q=0,r_0,\omega_s) & = r_0 \frac{(1+\frac{a}{b_s})}{(1-\frac{a}{b_s})} \; ,
 \end{align}
which agrees with the result in \cite{SVI1}.
Next, using the solutions of $r_0$ in (\ref{eq:r0}) and $X$ in (\ref{eq:Xsol}), from (\ref{upow1}) one can find the function $Y$ in
terms of $(a,Q,b,s,r_0)$.
Thus, we have successfully generalized the expressions of the Kerr black hole case in \cite{SVI1} to those of a Kerr-Newman black hole.
They become very
crucial, as seen later, to obtain an expression of the deflection angle of light rays in terms of the elliptic functions.

%\textcolor{blue}
{In the case of the Reissner-Nordstrom black holes ($a\rightarrow 0$), the expression in the  bracket of the deflection
angle~\eqref{eq:ODEdu/dphi} reduces to 1. Moreover, as discussed in \cite{CHAS}, the equation $B(u)=0$ has four roots as in the
Kerr-Newman case, in particular evaluating  $r_0$  at $r_c$ in (\ref{r_cNR})
gives $X=2M (\frac{2Mr_c}{Q^2}-1)$ in (\ref{eq:Xsol}), which is the solution of (\ref{eq:X}) in the limit of $a\rightarrow 0$.
 As such, the root of  $u$ in  (\ref{u4})  is then  $u_4=\frac{1}{r_c}$, and together with $u_2$ becomes double roots at
 $\frac{1}{r_c}$.  Here we have found four roots in a general $r_0$ for nonzero $Q$ and $a$, consistent with the findings in the
 literature by taking an appropriate limit  for  Kerr black holes ($a\neq 0, Q=0$) or a Reissner-Nordstrom black hole ($Q\neq 0, a=0$).}

To proceed, we first write the function in the  bracket of~\eqref{eq:ODEdu/dphi}
%can be further written
as
\begin{align}
 & \frac{1-2Mu(1-\omega_s)}{1-2Mu+(a^2+Q^2)u^2}+\frac{Q^2u^2(1-\omega_s)}{1-2Mu+(a^2+Q^2)u^2} \no\\
 & =\frac{C_+}{u_+-u}+\frac{C_-}{u_--u}+\frac{C_{Q+}\,u}
 {u_+-u}+\frac{C_{Q-}\,u}{u_--u}\;,
 \end{align}
where
 \begin{align}
 u_\pm=\frac{M\pm\sqrt{M^2-(a^2+Q^2)}}{a^2+Q^2} \, .
 \end{align}
Solving for $C_+$, $C_-$, $C_{Q+}$ and $C_{Q-}$, we obtain
 \begin{align}
 C_+ & =\frac{2M(1-\omega_s)\left(M+\sqrt{M^2-(a^2+Q^2)}\right)-(a^2+Q^2)}{2(a^2+Q^2)\sqrt{M^2-(a^2+Q^2)}} \, ,\no\\
 C_- & =\frac{(a^2+Q^2)-2M(1-\omega_s)\left(M-\sqrt{M^2-(a^2+Q^2)}\right)}{2(a^2+Q^2)\sqrt{M^2-(a^2+Q^2)}} \, ,\no\\
 C_{Q+} & =\frac{-Q^2(1-\omega_s)\left(M+\sqrt{M^2-(a^2+Q^2)}\right)}{2(a^2+Q^2)\sqrt{M^2-(a^2+Q^2)}} \, ,\no\\
 C_{Q-} & =\frac{Q^2(1-\omega_s)\left(M-\sqrt{M^2-(a^2+Q^2)}\right)}{2(a^2+Q^2)\sqrt{M^2-(a^2+Q^2)}} \, .
 \end{align}

 \begin{figure}[h]
 \centering
 \includegraphics[width=0.9\columnwidth=0.5]{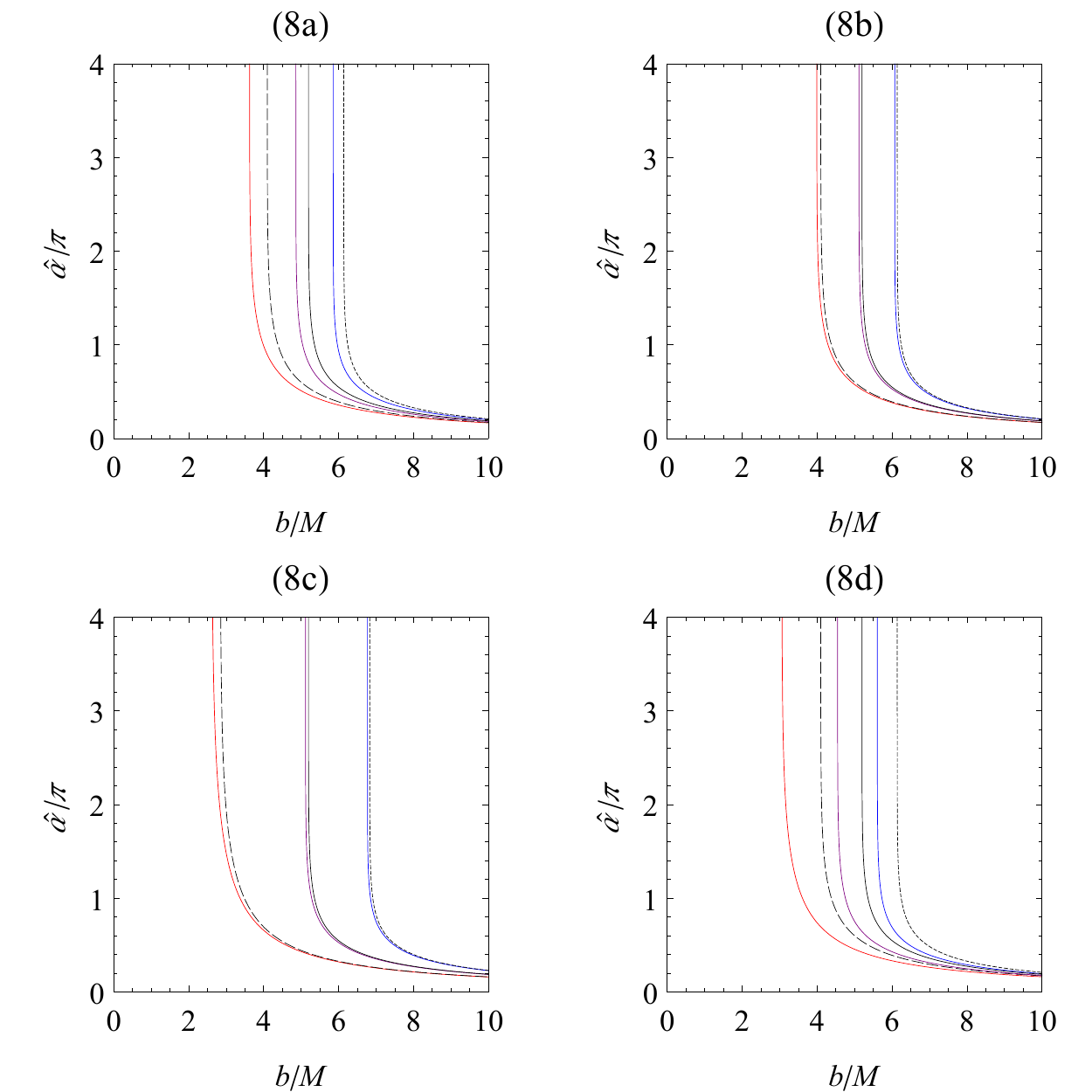}
 \caption{
 %\textcolor{blue}
  {
 Deflection angle as a function of impact parameter $b/M$ for four sets of parameters.
(a) $a/M=0.5$, $Q/M=0.6$; (b) $a/M=0.5$, $Q/M=0.3$; (c) $a/M=0.9$, $Q/M=0.3$; (d) $a/M=0.5$, $Q/M=0.8$. The plots show the results for
the Schwarzschild, Reissner-Nordstrom, Kerr and Kerr-Newman black holes for comparison.
%\textcolor{blue}
{The suppression of the bending
angle for light rays due to the charged black hole as compared with the corresponding black hole with $Q=0$
is found for the same  impact parameter $b$.}\label{alpha_b}
}
 %(2a) Kerr-Newman direct with $a/M=0.5$ and $Q=M=0.3$ (red), Kerr-Newman retrograde with $a/M=0.5$ and %$Q=M=0.3$ (blue), Kerr directwith $a/M=0.5$ (dashed), Kerr retrograde with $a/M=0.5$ and $Q=M=0.3$ (dotted), %Reissner-Nordstrom with $Q/M=0.3$ (purple), andSchwartzschild (black). The impact parameter is $b=8M$
 }
 %{Deflection angle as a function of impact parameter $b/M$ for Schwartzschild, kerr, and Kerr-Newman for comparison }}
 \label{fig:bsc_Q}
 \end{figure}

The integral form of deflection angle in~\eqref{eq:alpha} is then
 \begin{align} \label{alpha_kn}
 \hat{\alpha} & =-\pi+2\int_{0}^{1/r_0} \left(\frac{C_+}{u_+-u}+\frac{C_-}{u_--u}+\frac{C_{Q+}\,u}{u_+-u}+\frac{C_{Q-}\,u}
 {u_--u}\right)\frac{1}{\sqrt{B(u)}} \; \mathrm{d}u
 \end{align}
where $B(u)$ is written as a product of the roots in (\ref{B_fun}). The exact expression of bending angle consequently is given in terms
of elliptical integrals as \cite{GR2007}
\begin{align}
 \hat{\alpha} & =-\pi+\frac{4}{(1-\omega_s)\sqrt{Q^2(u_4-u_2)(u_3-u_1)}} \no\\
 &\quad
 \cdot\Bigg\{\frac{C_++C_{Q+}u_1}{u_+-u_1}\left[\Pi(n_+,k)-\Pi(n_+,\psi_0,k)\right]+\frac{C_-+C_{Q-}u_1}{u_--u_1}\left[\Pi(n_-,k)-\Pi(n_-,\psi_0,k)\right]
 \no\\
 &\qquad -\frac{C_++C_{Q+}u_4}{u_+-u_4}\left[\Pi(n_+,k)-\Pi(n_+,\psi_0,k)-K(k)+F(\psi_0,k)\right] \no\\
 &\qquad -\frac{C_-+C_{Q-}u_4}{u_--u_4}\left[\Pi(n_-,k)-\Pi(n_-,\psi_0,k)-K(k)+F(\psi_0,k)\right]\Bigg\} \;. \label{eq:6}
 \end{align}
In (\ref{eq:6}),
\begin{align}
 n_\pm &
 %=\frac{(u_2-u_1)\left[4M^2r_0-Q^2(X+2Mr_0u_\pm)\right]}{(u_\pm-u_1)\left[4M^2r_0-Q^2(X+2M)\right]}
 =\frac{u_2-u_1}{u_\pm-u_1}\left[1+\frac{2MQ^2(1-r_0u_\pm)}{4M^2r_0-Q^2(X+2M)}\right]\;, \no\\
 k^2 & =\frac{(Y+6M-X)\left[8M^2r_0-Q^2(Y-2M+3X)\right]}{4Y\left[4M^2r_0-Q^2(X+2M)\right]}\;, \no\\
 \psi_0 & =\arcsin{\sqrt{\frac{(Y+2M-X)\left[4M^2r_0-Q^2(X+2M)\right]}{(Y+6M-X)(4M^2r_0-Q^2X)}}}\;, \label{eq:7}
 \end{align}
%
%
% \begin{align}
% \hat{\alpha}& %=-\pi+\frac{4}{(1-\omega_s)}\sqrt{\frac{r_0}{Y\left[1-\frac{Q^2(X+2M)}{4M^2r_0}\right]}} \no\\
% &\quad\\cdot\Bigg\{\frac{C_++C_{Q+}u_1}{(u_+-u_1)}\left[\Pi(n_+,k)-\Pi(n_+,\psi_0,k)\right]+\frac{C_-+C_{Q-}u_1}{(u_--u_1)}\left[\Pi(n_-,k)-\Pi(n_-,\psi_0,k)\right]\no\\
% &\qquad\+\frac{Q^2\left(C_++C_{Q+}u_4\right)(Y+6M-X)}{n_+(u_+-u_1)\left[8M^2r_0-2Q^2(X+2M)\right]}\left[\Pi(n_+,k)-\Pi(n_+,\psi_0,k)-K(k)+F(\psi_0,k)\right]\no\\
% &\qquad\+\frac{Q^2\left(C_-+C_{Q-}u_4\right)(Y+6M-X)}{n_-(u_--u_1)\left[8M^2r_0-2Q^2(X+2M)\right]}\left[\Pi(n_-,k)-\Pi(n_-,\psi_0,k)-K(k)+F(\psi_0,k)\right]\Bigg\}
% \end{align}
%
% \begin{align}
% n_\pm & =\frac{(u_2-u_1)}{(u_\pm-u_1)}\left[1+\frac{2MQ^2(1-r_0u_\pm)}{4M^2r_0-Q^2(X+2M)}\right] \, ,\no\\
% k^2 & =\frac{\left[8M^2r_0-Q^2(Y-2M+3X)\right](Y+6M-X)}{4Y\left[4M^2r_0-Q^2(X+2M)\right]} \, ,\no\\
% \psi_0 & =\arcsin{\sqrt{\frac{(Y+2M-X)\left[4M^2r_0-Q^2(X+2M)\right]}{(Y+6M-X)(4M^2r_0-Q^2X)}}} %\, ,
% \end{align}
 and $\Pi(n_+,k)$ and $\Pi(n_+,\psi_0,k)$ are, respectively, the complete and the incomplete elliptic integrals of the third kind.
 Additionally,  $K(k)$, the complete elliptic integral of the first kind, and $F(\psi_0,k)$, the incomplete elliptic integral of the
 first kind, are also involved. In particular, the condition $0 \le k^2 \le 1$ has been checked numerically. This is one of the main results of the work.

\begin{figure}[h]
 \centering
 \includegraphics[width=0.9\columnwidth=0.6]{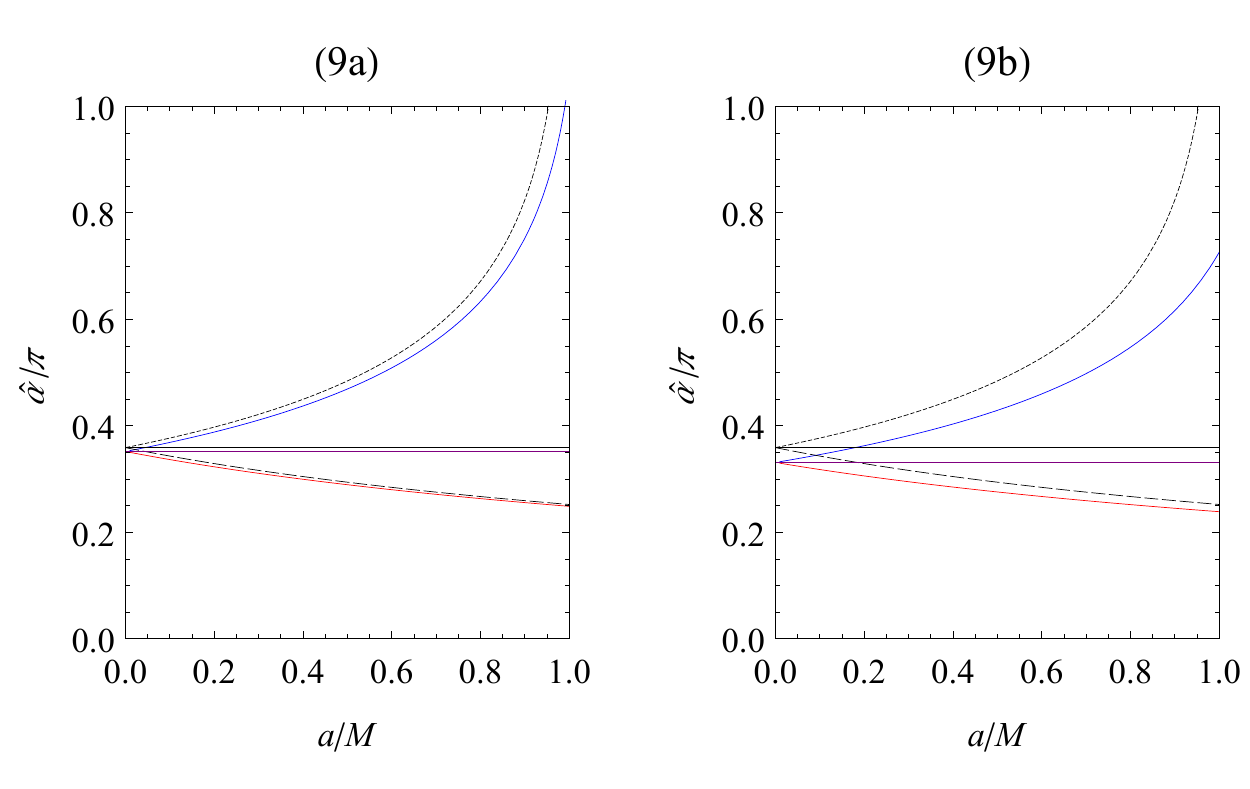}
 \caption{
 %\textcolor{blue}
 {Deflection angle as a function of spin parameter $a/M$ for (a) $Q/M=0.3$, and $b=7M$; (b) $Q/M=0.6$, and $b=7M$. The plots show the
 results for the Schwarzschild, Reissner-Nordstrom, Kerr and Kerr-Newman black holes for comparison.
 %\textcolor{blue}
 { The nonzero
 charge of the black hole  decreases the deflection angle  also due to the additional repulsive effects on the light rays for the fixed
 spin of black hole $a$ and the given impact parameter of the light ray $b$.}}
 \label{alpha_a}}
 \end{figure}

\begin{figure}[h]
 \centering
 \includegraphics[width=0.9\columnwidth=0.6]{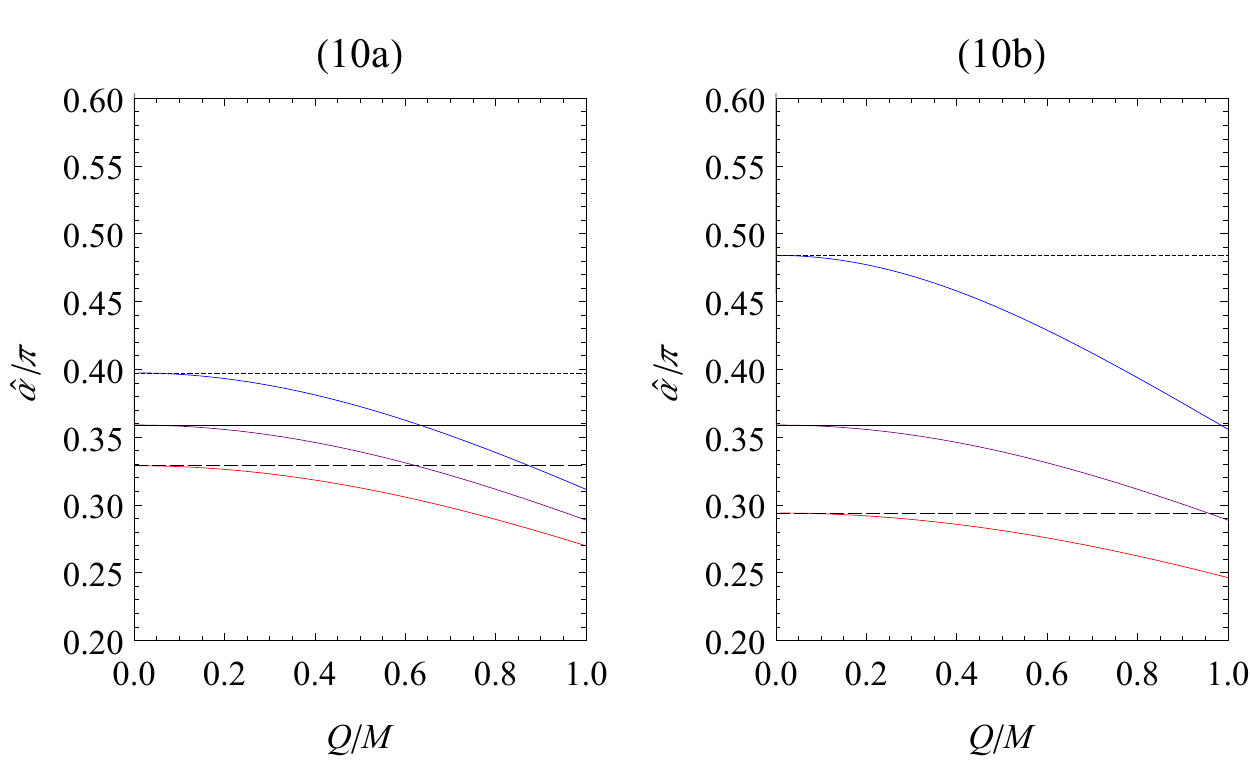}
 \caption{
 %\textcolor{blue}
 {Deflection angle as a function of charge $Q/M$ for (a) $a/M=0.2$, and $b=7M$; (b) $a/M=0.5$, and $b=7M$. The plots show  the results
 for the Schwarzschild, Reissner-Nordstrom, Kerr and Kerr-Newman black holes for comparison.
 %\textcolor{blue}
 {The deflection angle
 decreases as the charge $Q$ of the black hole increases for a given impact parameter $b$.}}\label{alpha_Q}}
 \end{figure}

It is quite straightforward to find that in the $Q=0$  limit, the whole expression of (\ref{eq:6})
reduces to that of the Kerr black hole in \cite{SVI1}.
%\textcolor{blue}
{Additionally, the deflection angle  (\ref{eq:6}) in the limit of
$a\rightarrow 0$ gives the closed-form expression of the bending angle of light rays due to  the Reissner-Nordstrom black holes}.
Numerical studies in Fig.\ref{alpha_b}   reproduce the deflection angle  of light rays due to the  Schwarzschild $(Q\rightarrow 0, a
\rightarrow 0)$ and Kerr black holes $(Q\rightarrow 0)$ as a function of the impact parameter $b$ in \cite{SVI1}.
However, in the Kerr-Newman case, one of the remarkable results is the suppression of the bending angle compared with the Kerr case
due to the  repulsive effects from the black hole's charge to the light rays for a fixed impact parameter in both direct and retrograde
motions. These effects can be realized in  Fig.\ref{alpha_a}, and also in particular in Fig.\ref{alpha_Q} where the deflection angle
$\hat\alpha$ decreases as the charge $Q$ of the black hole increases.
This will result in the modification of the apparent shadows from the Kerr case as it has been  seen in \cite{VRI4}.
%\textcolor{blue}
{The shadow of the  black hole is perceived by an observer at spatial infinity at different polar positions. Based upon
our results, it is anticipated that the larger charge $Q$ with a fixed black hole spin and the polar positions of the observers leads to
the smaller boundary of the shadow of the Kerr-Newman black holes that will be further explored in our future work. }
Our approach gives a closed-form expression of the deflection angle of the light rays due to the Kerr-Newman black hole that certainly
generalizes
the Kerr case in \cite{SVI1}, and also explores the effect from the charge of the black hole to the light bending  with an appropriate
``effective potential" defined from the radial motion of a light ray.

\section{Summary and outlook}
In summary, the dynamics of light rays traveling around a Kerr-Newman black hole is studied with emphasis on how the  charge of  black holes influences their trajectories. We first {examine} the innermost  circular orbits on the equatorial plane. It is found that the presence of the charge of the black hole results in the  ``effective potential" which is defined from the equation of motion of light rays  along the radial direction,  with stronger repulsive effects as compared with the Kerr back hole.
As a result, the radius of the innermost circular motion of light rays decreases as charge $Q$ of the black hole increases for both direct and retrograde motions, which certainly gives a direct consequence on constructing the apparent shape of a charged rotating black hole.
We  then  derive the closed-form expression  of the deflection angle in terms of elliptic integrals by generalizing the work of
  \cite{SVI1}.
The nonzero charge of the black hole  decreases the deflection angle  also due to the additional repulsive effects on the
  light rays given by the ``effective potential" of the photon radial motion.
{In the next step},  we will study this analytical solution
in the form of series expansion in the strong and weak deflection regimes, and compare with  the works of ~\cite{BAR,SVI3}  for the Kerr
case.
 The accurate and efficient approximation schemes can be an attractive alternative to compute the involved elliptical integrals in black
 hole simulations.

\begin{acknowledgments}
This work was supported in part by the Ministry of Science and Technology, under Grant
No. 108-2112-M-259-002.
\end{acknowledgments}

\end{document}